# Automated Whole Slide Imaging for Label-Free Histology using Photon Absorption Remote Sensing Microscopy

James E.D. Tweel[1,2], Benjamin R. Ecclestone[1,2], Marian Boktor[1], Deepak Dinakaran[2], John R. Mackey[2], Parsin Haji Reza[1,*]

[1]PhotoMedicine Labs, University of Waterloo, 200 University Ave W, Waterloo, ON N2L 3G1
[2]illumiSonics Inc., 22 King Street South, Suite 300, Waterloo, ON, N2J 1N8
*Corresponding author: phajireza@uwaterloo.ca

**Abstract** – The field of histology relies heavily on antiquated tissue processing and staining techniques that limit the efficiency of pathologic diagnoses of cancer and other diseases. Current staining and advanced labeling methods are often destructive and mutually incompatible, requiring new tissue sections for each stain. This prolongs the diagnostic process and depletes valuable biopsy samples. In this study, we present an alternative label-free histology platform using the first transmission-mode Photon Absorption Remote Sensing microscope. Optimized for automated whole slide scanning of unstained tissue samples, the system provides slide images at magnifications up to 40x that are fully compatible with existing digital pathology tools. The scans capture high quality and high-resolution images with subcellular diagnostic detail. After imaging, samples remain suitable for histochemical, immunohistochemical, and other staining techniques. Scattering and absorption (radiative and non-radiative) contrasts are shown in whole slide images of malignant human breast and skin tissues samples. Clinically relevant features are highlighted, and close correspondence and analogous contrast is demonstrated with one-to-one gold standard H&E stained images. Our previously reported pix2pix virtual staining model is applied to an entire whole slide image, showcasing the potential of this approach in whole slide label-free H&E emulation. This work is a critical advance for integrating label-free optical methods into standard histopathology workflows, both enhancing diagnostic efficiency, and broadening the number of stains that can be applied while preserving valuable tissue samples.

## I. INTRODUCTION

HISTOLOGY is the study of microscopic cell structure and function and is critical in understanding the biological processes that underlie health and disease [1]. Histopathologists use a variety of staining techniques on tissue samples to highlight cell structure and composition, distinguish tissue types, and localize different molecules such as proteins, lipids, and carbohydrates. The contrast provided by these stains can show abnormalities in the tissue, indicating the presence, nature, and extent of a diseases like cancer. Prior to staining, fresh tissue samples are formalin fixed, dehydrated, paraffin embedding and thinly sliced (~5µm) onto microscope slides [1], [2]. Tissue sections are then stained and imaged for microscopic observation, using a brightfield or fluorescent microscope. The gold standard stain in histology, widely used in cancer diagnosis, is hematoxylin and eosin (H&E). It stains the chromatin in the nuclei purple with hematoxylin, while staining the cytoplasm and extracellular structures pink with eosin [2]. There are also specialized stains, such as Masson's trichrome and periodic acid Schiff, capable of targeting more specific cellular or tissue structures [3]. In addition, more advanced techniques such as immunohistochemical staining and *in-situ* hybridization are available for localization of specific proteins and DNA/RNA sequences, respectively [3]. However, these selective stains and advanced techniques are often incompatible and tend to interfere [4], [5]. In practice, new tissue sections are required for each stain or technique applied. However, tissue processing steps are time-consuming, resource intensive, expensive, and require trained histotechnologists [6]. If multiple sequential stains are required for complete diagnosis, treatment may be delayed, and outcomes may worsen. Furthermore, tissue processing and staining are destructive. Cutting through multiple tissue sections for each stain can quickly deplete valuable biopsy samples and increase the likelihood of needing a repeat biopsy from patients, further prolonging the process.

Advancements in digital pathology, such as whole slide imaging scanners and machine learning algorithms, have improved standard histopathology workflows as well as accuracy and efficiency of diagnosis [7], [8]. However, reliance on century-old tissue processing and staining techniques limits progress and overall impact. Label-free imaging platforms are needed to overcome this fundamental limitation. Such techniques exploit the photophysical processes of biomolecules to visualize inherent contrasts within a sample without the use of chemical labels (e.g., stains or fluorescent dyes) [9]. This opens the possibility of simultaneous visualization of multiple histological or immunobiological-like stains from a single unaltered tissue section, providing a more complete diagnostic overview and reducing preparation time, resources and required expertise. As well, valuable biopsy samples remain preserved and available for redundant or alternate screening procedures.

Broadly speaking, current label-free optical methods use scattering and absorption processes to provide contrast in biological specimens. Scattering techniques such as optical coherence tomography (OCT) have been studied as a potential tool for label-free tissue histology, with previous research demonstrating its ability to achieve cellular scale resolutions in volumetric acquisitions of bulk tissue specimens [10]–[13].

Similarly, quantitative phase imaging (QPI) methods have been investigated for label-free cell and tissue imaging, diagnostics and virtual histology [14]–[17]. QPI is able to acquire refractive index distributions across a tissue sample and provide nanoscale visualizations of tissue structures [14]. However, QPI strictly relies on transparent samples, limiting its ability in bulk tissue imaging. Furthermore, both QPI and OCT rely solely on tissue scattering properties which lack biomolecular specificity [10], [15]. As a result, their diagnostic utility largely depends on interpreting morphological features alone.

On the other hand, absorption-based microscopy techniques are better able to target biomolecules given their unique absorption spectra [18], [19]. In recent years, autofluorescence-based microscopy techniques have shown potential in label-free histology [20]–[23]. Many biologically relevant molecules and tissue structures become autofluorescent when excited by an appropriate wavelength, typically in the ultraviolet (UV) or shorter wavelength visible range [24]. The emission from these endogenous fluorophores contains valuable information which correlates to structural and functional characteristics of biological specimen [20], [24], [25]. However, while these contrasts map well to cytoplasm and extracellular matrix structures, chromatin (DNA, RNA) has relatively low quantum yield [26]. As a result, important nuclear contrast is generally missing with such techniques. Other non-linear microscopy techniques, including two-photon and three-photon autofluorescence, second and third harmonic generation, and Raman scattering modalities have been used independently [27]–[31] and in multimodal configurations [32], [33] to provide extra dimensionality in radiative absorption contrast for use in stain-free histology.

Non-radiative optical absorption contrast has also been realized in photothermal [34], [35] and photoacoustic (PA) [36]–[38] imaging techniques for label-free histology-like visualizations. In particular, optical-resolution photoacoustic microscopy (OR-PAM) methods have shown high-resolution visualizations of nuclear and cytoplasmic structures [36]–[38]. In addition, photoacoustic remote sensing, a non-contact optical absorption-based method, was first used for label-free nuclear imaging in unstained tissues by Haven et al. [39] and Abassi et al. [40] in 2019. It has also been used to image a variety of biomolecules label-free including hemoglobin [41]–[45], cytochromes [46], [47], DNA/RNA and lipids [48], [49]. Recently, Ecclestone et al. [50] introduced a new generation of photoacoustic remote sensing, now called photon absorption remote sensing (PARS). PARS uniquely captures contrasts afforded by both the radiative and non-radiative relaxation processes following a single absorption event from a pulsed excitation laser. Following excitation, all optical emissions from the radiative relaxation process are broadly captured while heat and pressure, from the non-radiative process, modulate the sample's local optical properties. Non-radiative contrast is derived from the resulting intensity modulations (photothermal and photoacoustic signals) of a secondary co-focused detection beam's backscattered light [50]. Prior to excitation, this backscattered light is used to capture scattering contrast of the sample. PARS paired with UV excitation is able to capture cell and nuclear structures (non-radiative) as well collagen, cytoplasm, and connective tissue details (radiative channel). Recently, using generative machine learning models, these contrasts have combined to virtually stain samples with H&E [51].

Here we have developed the first transmission mode PARS microscope optimized for label-free digital histology. This development includes key optic refinements for superior image quality from previous reports and image resolutions necessary for standard 40x magnification viewing. We present an automated whole slide scanning workflow for the PARS system which includes a novel contrast leveling algorithm for whole slide stitching. The label-free whole slide images shown here allow for high-resolution examination of tissue structures at any magnification level up to 40x. This provides pathologists with the required magnification range to analyze entire tissue samples for informed and accurate diagnosis. The label-free images provide contrasts analogous to gold standard H&E stained slides and are viewable using the same or similar existing digital pathology tools. We also show, for the first time, virtual H&E staining of an entire whole slide PARS image. The unstained tissue sections imaged with this device remain unaltered and available for subsequent standard histochemical staining or auxiliary techniques. Here, both the PARS and H&E whole slide images are acquired from the same tissue section in malignant human breast and skin samples. Important nuclear and extracellular structures are compared between these one-to-one whole slide image datasets.

## II. METHODS

### A. PARS Histology System Architecture

The PARS histology imaging system architecture is shown in Figure 1. For the excitation laser, a 50KHz 400ps pulsed 266nm UV laser (Wedge XF 266, RPMC) was chosen. A $CaF_2$ prism (Prism: PS862, Thorlabs) was used to remove residual 532nm light from the source and send it to a beam trap (BT: BT610, Thorlabs). The UV beam is then expanded (VBE: BE03-266, Thorlabs) and combined with the detection pathway using a long-pass dichroic mirror (DM: 37-721, Edmund optics). For the detection path, a continuous wave 405nm laser (OBIS-LS 405, Coherent) was chosen. The fiber-coupled sourced is collimated (Col: C40APC-A, Thorlabs) and then directed through the dichroic mirror to join the UV beam. The co-aligned detection and excitation beams are then focused onto the sample with a 0.42 numerical aperture (NA) UV objective lens (OL1: NPAL-50-UV-YSTF, OptoSigma). The transmitted detection light and forward emissions from the radiative relaxation process are collected using a 0.7 NA objective lens (OL2: 278-806-3, Mitutoyo) and then separated with a 405nm notch filter (SF: NF405-13, Thorlabs). The detection intensity modulations and radiative emissions are then each recorded on an avalanche photodiode (PD: APD130A2) following condenser lenses (Cond: ACL25416U-A, Thorlabs). Prior to excitation and detection co-alignment, beam samplers (BSF10-UV/A, Thorlabs) are used to sample a portion of both the detection power and excitation pulse energy for reference and post-acquisition correction purposes.

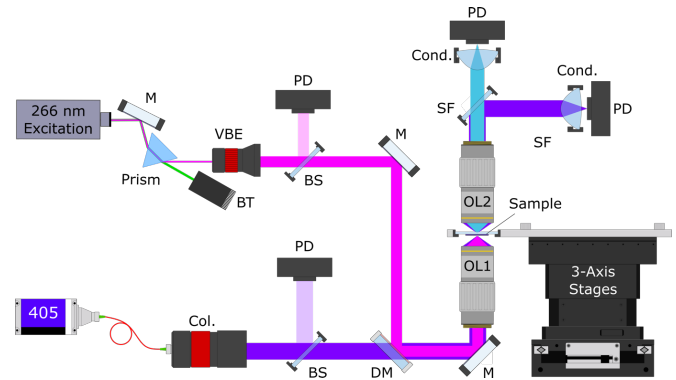

Figure 1: Simplified PARS histology optical architecture. Component labels are defined as follows: mirror, *M*; dichroic mirror, *DM*; variable beam expander, *VBE*; collimator, *Col.*; condenser lens, *Cond.*; spectral filter, *SF*; beam sampler, *BS*; photodiode, PD; objective lens, OL.

### B. Whole Slide Scanning Workflow

The PARS histology system first captures a preview image of the whole slide to allow the user to trace a border around the entire sample or draw a smaller region of interest (ROI). These ROI coordinates are mapped to mechanical stage positions to inform the scanning algorithm of the tissue boundaries. The optimal focus plane must be maintained across the entire sample in order to acquire consistent high quality raw data. To do so, the selected area is split into smaller (typically ~0.5mm²) square sections ($S_n$) as shown in Figure 2. Sections are then scanned in sequence at their respective optimal focus plane by adjusting the axial (z) stage position. With this approach, the system is able to compensate for variances in the sample's surface morphology and is robust to any tilt in the sample and scanning plane. Some sections positioned at the edge of the sample may cover more glass area than tissue area. This may cause the optimal focus to skew towards particles on the glass surface rather than tissue layer. As such, these section are flagged (blue in Figure 2(a)) and will be scanned at the optimal focus of the closest adjacent section with sufficient tissue coverage (red in Figure 2(a)).

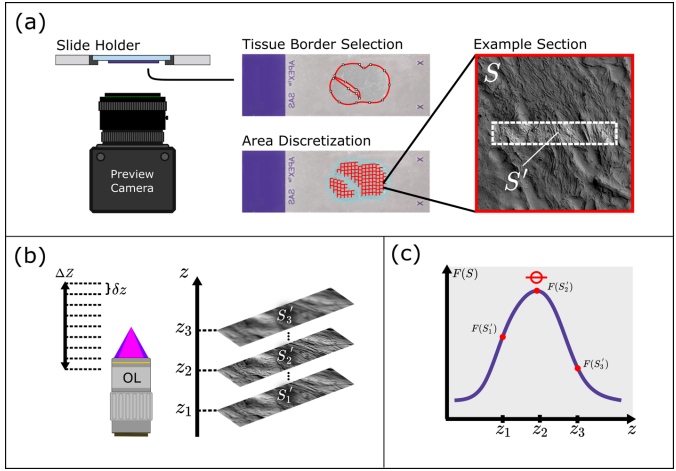

Figure 2: Overview of the PARS whole slide scanning workflow. (a) A preview camera is used to capture an image of the slide so the tissue border can be selected, and its area split into small sections, each to be independently scanned. (b) Before imaging $S_n$, scattering images of subset $S'_n$ are taken across a depth $\Delta Z$ at a step size $\delta z$. (c) A relative focus metric is plotted for each scattering image from (b), and an optimal focus metric is determined.

### C. Focus Criterion Function and Autofocus Algorithm

To find the optimal plane of focus for each section, a small sampling of the section area ($S'_n$) is imaged multiple times across a depth of $\Delta Z$ at step sizes of $\delta z$, as shown in Figure 2(b). For these subsections, the detection beam alone is sufficient for finding peak focus for all image channels (scattering, radiative and non-radiative). The Tenengrad focus criterion function is then used to evaluate the relative sharpness of the scattering images acquired across depth. The Tenengrad function was chosen because it provides a sharp peak for fine focusing and is relatively noise robust [52]. It first computes the horizontal ($G_x$) and vertical gradients ($G_y$) of an input image ($I$) at point ($x, y$) using the Sobel kernels:

$$G_x = \begin{bmatrix} 1 & 0 & -1 \\ 2 & 0 & -2 \\ 1 & 0 & -1 \end{bmatrix} * I \quad \text{and} \quad G_y = \begin{bmatrix} 1 & 2 & 1 \\ 0 & 0 & 0 \\ -1 & -2 & -1 \end{bmatrix} * I \quad (1)$$

From here the gradient magnitude, $G(x,y)$, is computed:

$$G(x,y) = \sqrt{G_x^2(x,y) + G_y^2(x,y)} \quad (2)$$

The focus criterion function, $f(I)$, is then evaluated as the sum of squared gradient magnitude values which exceed a given threshold T:

$$F(I) = \sum_x \sum_y [G(x,y)]^2 \quad \forall \; G(x,y) > T \quad (3)$$

The Tenengrad function quantifies the magnitude of the edges present in the scene. Across depth, it will produce a plot similar to Figure 2(c), with a single peak which monotonically decreases as the sample moves above or below this point. At the beginning of the whole slide scan, a course focus search is deployed over a large range to determine a rough starting point for the slide. Following this, subsequent section foci are found by quadratically interpolating the evaluated depth stack over a narrower range ($\Delta Z$) with a center point equal to the optimal focus point of the closest neighbouring section.

### D. Section Scanning and Image Reconstruction

During imaging, mechanical stages continually move the sample over the objective lens in an 's' or 'snake'-like scan pattern. Sample points are spaced out laterally to achieve the desired 40x equivalent magnification (~250nm/pixel). At each excitation event, a high-speed digitizer (CSE1442, RZE-004-200, Gage Applied) captures the stage position information as well as ~500ns of time resolved data from each system photodiode. These signals are then compressed into single amplitude values for the non-radiative, radiative, and scattering image channels. In addition, the 266nm excitation pulse energy and 405nm detection power are also simultaneously collected at this location. To form the non-radiative, radiative, and scattering images, the raw data points are arranged as pixels on a cartesian grid based on their stage position. The radiative and non-radiative images are then power corrected using the 266nm pulse energy measurement. Similarly, the scattering image is power corrected using the 405nm detection power measurement.

### E. Whole Slide Stitching and Contrast Leveling Algorithm

After all sections are scanned, they can be stitched together into a whole slide image using their relative stage positions. However, prior to stitching, a contrast and brightness leveling algorithm is needed to blend the shading differences between adjacent sections. The algorithm is two part, consisting of a bulk leveling stage (Figure 3(a)) and gradient leveling stage (Figure 3(b)). Both parts use a small amount of overlap (~100 pixels per direction) between sections to help with the blending. The bulk leveling algorithm brings all sections into the same overall brightness and contrast range while the gradient leveling algorithm blends the sharp transitions between sections due mismatched gradients of illuminations. In bulk leveling, the inner and outer overlap areas are used to determine a transform ($T_{\mu\sigma_{i/o}}$) to level the inner section. This transform is based on the mean and standard deviation of the inner ($\mu_i$, $\sigma_i$) and outer overlap pixels ($\mu_o$, $\sigma_o$).

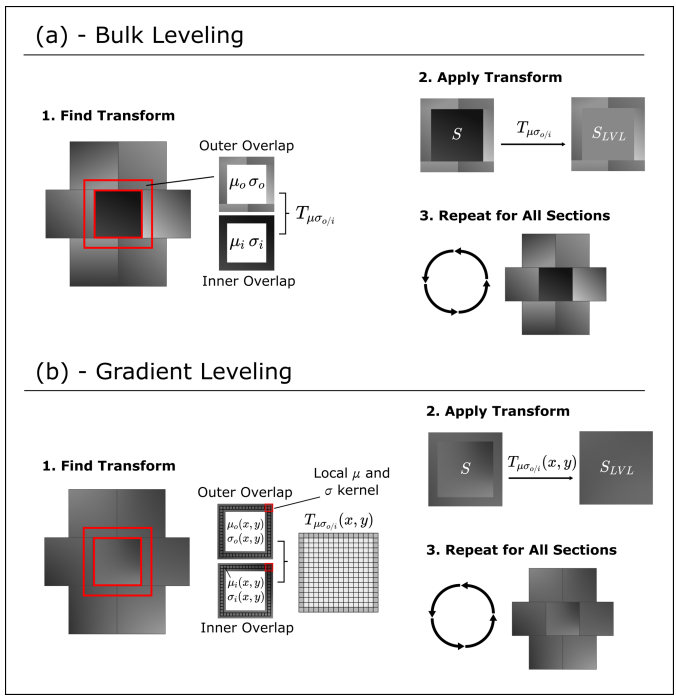

Figure 3: Illustration of the two part algorithm used to level and blend the contrast and shading between sections. (a) Bulk leveling algorithm aims to bring all sections into the same general contrast range. Here a single transform ($T_{\mu\sigma_{o/i}}$) for the entire inner section is derived based on the overlap area between the inner and outer sections. (b) A gradient leveling algorithm aimed at continuously blending the boundaries between neighbouring sections. Here a point transform ($T_{\mu\sigma_{o/i}}(x,y)$) for the inner section is derived based on a comparison between the local statistics of the grayscale values of inner and outer overlap regions.

$$S_{LVL} = T_{\mu\sigma_{i/o}}(S) = \frac{\sigma_o}{\sigma_i} * (S - \mu_i) + \mu_o \qquad (4)$$

Equation 4 scales the inner section's pixel intensity spread to better match the outer section based on the ratio of the outer and inner overlap standard deviations. It then shifts the average pixel intensity of the inner section ($S$) to the average pixel intensity of the outer section overlap. This is then repeated for every section in the whole slide image before moving on to the gradient leveling algorithm. For gradient leveling, a kernel with a size equal to the overlap, computes a local mean and standard deviation value at each point along the inner and outer overlaps. Values in non-overlap locations are interpolated using values from surrounding overlap areas. These local statistic values are used to create a point transform, $T_{\mu\sigma_{i/o}}(x,y)$, which has the same form as equation 4 but varies at each pixel location. The transform values are then gaussian blurred and applied to the inner section. This is then repeated for every section in the whole slide image. In some cases, the bulk or gradient leveling algorithm may be repeated more than once until the mosaic artifacts are corrected.

### F. Sample Preparation

In this study, a variety of unstained paraffin embedded human skin and breast tissue sections were imaged on glass microscope slides. The tissues were first placed in formalin fixative solution for 24 to 48 hours within 20 minutes of excision. After fixation, the tissues were dehydrated following exposure to ethanol. Xylene was then used to remove the remaining ethanol and residual fats to prepare the tissue for wax penetration. Tissues were then embedded with paraffin wax to create formalin fixed paraffin embedded tissue (FFPE) blocks. Using a microtome, thin tissue sections (~4-5 μm) were sliced from the FFPE block surface and placed onto glass microscope slides. The tissue slides were then baked at 60ºC for ~60 minutes to remove excess paraffin prior to imaging with the PARS system. Following PARS imaging the exact same tissue sections were stained with H&E and imaged at 40x magnification using a standard brightfield microscope (MorphoLens 1). This provided a direct one to one comparison between the PARS image data and the gold standard H&E stain.

Tissues were provided by clinical collaborators at the Cross-Cancer Institute (Edmonton, Alberta, Canada) from anonymous patient donors with all patient identification removed from the samples. Samples were archival tissues no longer required for patient diagnostic and thus patient consent was waived by the ethics committee. No information was provided to the researchers about the patient identity. Samples were collected under protocols approved by the Research Ethics Board of Alberta (Protocol ID: HREBA.CC-18-0277) and the University of Waterloo Health Research Ethics Committee (Photoacoustic Remote Sensing (PARS) Microscopy of Surgical Resection, Needle Biopsy, and Pathology Specimens; Protocol ID: 40275). All human tissue experiments were conducted in accordance with the government of Canada guidelines and regulations, such as "Ethical Conduct for Research Involving Humans (TCPS 2)".

### III. RESULTS AND DISCUSSION

### A. System Improvements and Power Correction

The reflection based PARS microscope used in previous reports [50], [51] has been modified into a transmission based architecture for scanning thin transmissible samples. In this configuration, the stronger forward scattering allows for lower detection power while maintaining the similar recovery of the modulated detection signal from the non-radiative relaxation process. Transmission mode is also less sensitive to surface roughness and allows for a simplified architecture, as there is no requirement for separation of the detection back path. Furthermore, separation of the 266nm UV excitation pulse is not required in the forward path as it is absorbed by the borosilicate glass slide. Optical section is also not required for slide scanning because tissue sections are already thinly cut (~5µm). Therefore, the confocal architecture in the detection path has been removed. As such, forward scattered light is collected from the entire depth of focus of the detection spot. This is more similar to what happens when tissue sections are captured under brightfield illumination. For the radiative relaxation pathway, a 405nm notch filter has been added to spectrally separate the forward detection path from the radiative emission spectrum. Another important addition to the system were the beam samplers used to record the excitation pulse energy and detection power as the samples were point scanned. Figure 4 highlights improvements in image quality for the non-radiative, radiative, and scattering channels after power compensating the images. In the non-radiative and radiative image channels, each pixel is derived from a single excitation event. As such, pulse energy variability results in varying absorption which manifests as a speckle-type noise on a pixel or sub-resolution level. In the case of the scattering image, which are collected with a continuous wave laser, power instabilities result in slow varying intensity streaks as the sample is scanned. The non-radiative and radiative resolutions here are measured at ~350nm (FWHM), which roughly corresponds to standard 40x magnification in standard histopathology [53]. By using the collected excitation pulse energy and detection power measurements these artifacts can largely be removed and the 40x magnification is well maintained. Furthermore, power compensation in the detection scattering image allowed for more robust focus finding algorithm, which was important for the automated whole slide scanning reported here.

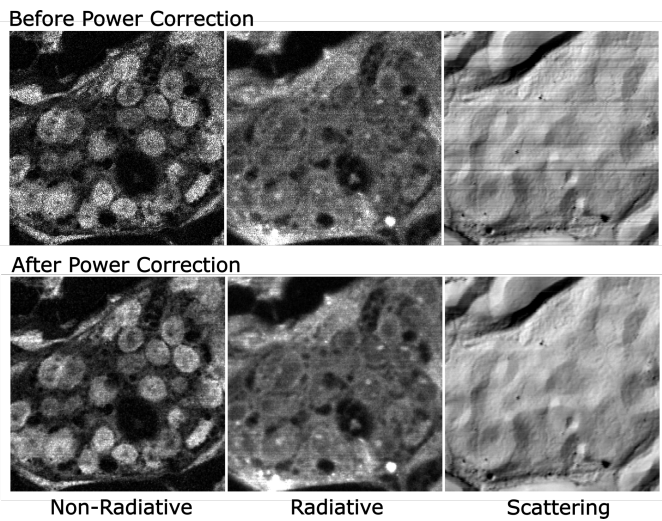

Figure 4: Non-radiative, radiative, and scattering label-free PARS image contrasts before and after correction from the measured reference power, demonstrated on a cluster of cell nuclei.

## B. Sample Autofocusing and Whole Slide Stitching

In the ideal case, the highest signal to noise ratio (SNR) and in-focus non-radiative and radiative acquisitions occurs at the optimal focus point of the detection beam. This way, the detection beam alone can be relied on for focus finding and the sample need not be exposed multiple times to UV excitation. To accomplish this, the optimal scattering focus is first determined using the methods described in section II.A. At this peak focus, the excitation spot is moved to the optimal axial position where the largest refractive index modulation is observed. Following this axial alignment strategy, a tight correspondence between the detection focus and peak SNR of the non-radiative channel can be seen in Figure 5(a). Here a series of scattering images are acquired across a +/- 35µm depth range at 1µm steps. The SNR is also evaluated for a series of non-radiative images captured over a +/18 µm range at 1µm steps. Figure 5(b) shows a stack of non-radiative images captured at and on either side (+/-1.5µm and 3µm) of the peak detection focus. Clear visual degradation of the non-radiative channel can be seen with scans taken on either side of the focus spot, which corresponds well with the sharp SNR drop on either side of the focus. Overall, Figure 5 emphasizes and confirms the importance of maintaining focus across the entire sample. For this reason, the whole slide images are captured in discrete in-focus sections which are stitched back together. However, the final stitched whole slide image often contains tiling artifacts between the edges of the sections. This is due to non-level illumination across sections caused by variances in tissue surface morphology. These artifacts can be corrected using the contrast leveling algorithm discussed in section II.E, allowing for better image interpretation. Figure 6 shows whole slide scattering images of a small breast tissue section (~3.5 x 1.2mm$^2$) before and after the bulk and gradient leveling algorithms. The whole slide image shown here, and all subsequent images, were captured with a section size of 0.5x0.5mm$^2$ and an overlap of 40µm. Figure 6(a) provides an example of the raw (pre-leveled) stitched whole slide image. In this raw image, clear brightness variations can be seen between sections and the uneven shading at section boundaries results in sharp stitching artifacts. After bulk leveling the whole slide, the obvious brightness variations between sections have largely been removed and all sections have been normalized to the same brightness range (Figure 6(b)). However, stitching artifacts are still visible due to mismatched shading gradients between sections. After applying the gradient leveling algorithm, these sharp section transitions have been corrected giving the impression of a homogenous whole slide image (Figure 6(c)). Overall, leveling the whole slide image is important for improving image appearance, diagnostic value, and is a critical preprocessing step for virtual staining.

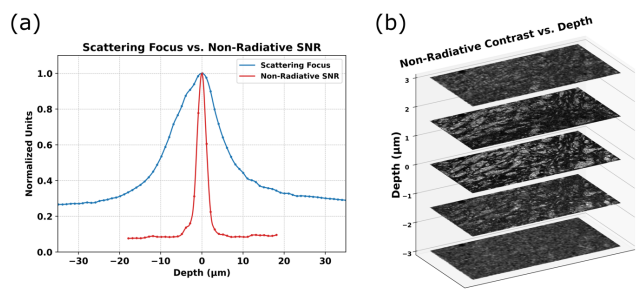

Figure 5: Relationship between scattering focus and the non-radiative image SNR. (a) Plot of the scattering focus across depth (blue) compared with non-radiative images SNR across depth (red). (b) Visualizations of non-radiative contrast quality across a +/- 3 µm depth.

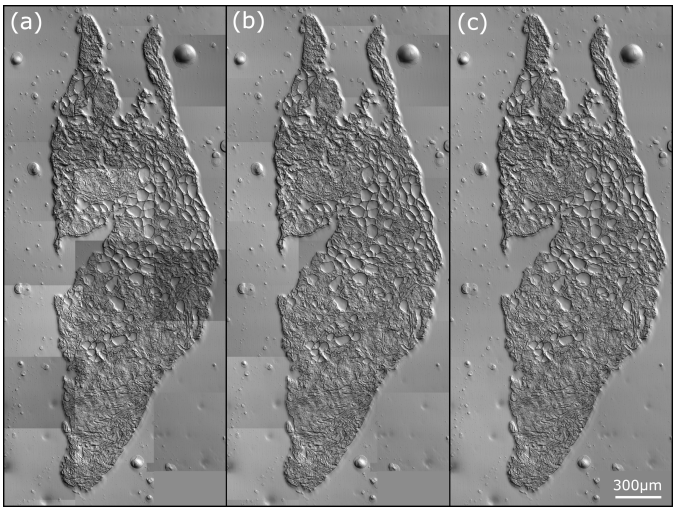

Figure 6: Whole slide scattering image at each stage of the brightness leveling algorithm. (a) Shows the raw stitched scattering image before leveling sections. (b) Shows the stitched scattering image after bulk leveling. (c) Shows the stitched scattering image after both bulk leveling and gradient leveling.

### C. Label-Free Whole Slide Images

Often pathologists will first analyze whole slide samples at low magnifications to get an overview of tissue structure, cell distribution and determine areas of interest. It provides the necessary context for the detailed examination of specific tissue structures at high magnification. It is therefore important to provide pathologists with entire whole slide images. Figure 7 shows an example whole slide image of a breast core needle biopsy with invasive ductal carcinoma. A heat map of the sample's relative focus positions is shown in Figure 7(a), indicating the topology of the sample's surface. Each section is outlined in white and has an optimal focus assigned to its center. All other focus values are interpolated between sections. Figure 7(b) shows zoomed-out whole slide images of the non-radiative and radiative channels while Figure 7(c) shows several higher magnification sections taken across the length of the tissue. The non-radiative channel predominantly highlights nuclear contrast while the radiative channel shows complimentary contrast of collagen, elastin structures and features of the extracellular matrix. The non-radiative channel shows a high degree of variance in nuclear size, shape, and appearance, which are indicative of neoplastic features often used in the evaluation of a cancer specimen. The non-radiative channel also highlights irregular loosely arranged glandular structures with a high degree of disorganisation, which are important features in grading malignancies (Figure 7(c) blue box). The abnormal cells are seen invading and infiltrating into the surrounding stroma and adipose tissue (Figure 7(c) red box). These visualizations, provided by the non-radiative channel, help pathologists determine the aggressiveness of the cancer that guides management recommendations. Furthermore, at high magnifications, subnuclear details such as nucleoli and chromatin/chromatid arrangement are seen in the radiative channel (Figure 7(c) blue box). Such structures help show the degree of nuclear aberration and cell division (mitotic figures) in the tissue, which are important for cancer diagnosis. Necrosis of breast tissue can also be seen in the radiative channel, inside a region of neoplastic cells, indicating high grade disease (Figure 7(c), green box). Fibroblasts, which contribute to the creation of connective tissue, are also seen in the non-radiative channel. In the corresponding radiative image, these fibroblasts are embedded in long diagonal strands of connective tissue (Figure 7(c) purple box).

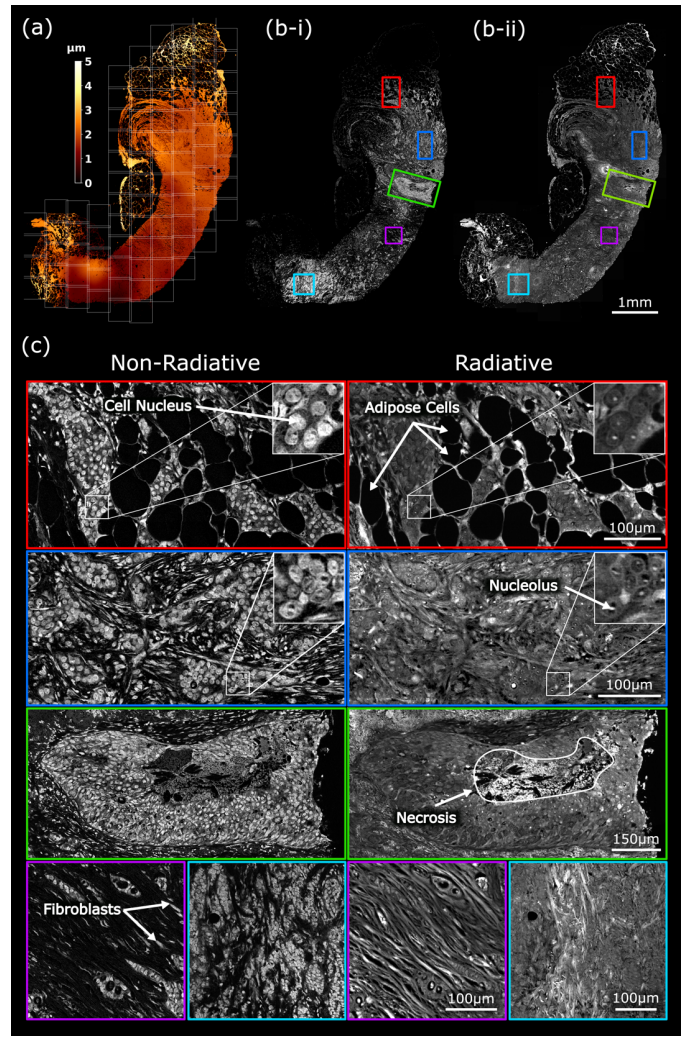

Figure 7: Stitched and leveled PARS non-radiative and radiative whole slide images of malignant human breast tissue. (a) Shows a heat map of the relative focus positions across the sample. (b-i) and (b-ii) Show whole slide images of both non-radiative and radiative channels, respectively. (c) Shows the non-radiative and radiative contrasts at high magnification. Sections are taken from across the sample and show clinically relevant features.

It's clear these label-free contrasts provide visualizations of clinically relevant features at both low and high magnifications. To further demonstrate their utility, these contrasts were combined into a single total absorption (TA) coloured image and compared against gold standard H&E staining. This one-to-one correspondence is shown in Figure 8 for the same high magnification sections shown previously. The TA image here has the radiative channel in blue and non-radiative in red. In general, the non-radiative channel provides analogous contrast to the hematoxylin stain while the radiative channel provides analogous contrast to the eosin stain. Correspondence of the nuclear structure, shape, and size and subnuclear details is clearly seen between the TA and H&E images. Much like the combined hematoxylin and eosin stains, the combined radiative and non-radiative contrasts allow clear differentiation of various tissue types and cell structures. The potential for label-free H&E emulation is clear. The H&E images were reviewed with clinicians and found to be of diagnostic quality with no artifacts apparent from the prior PARS scanning.

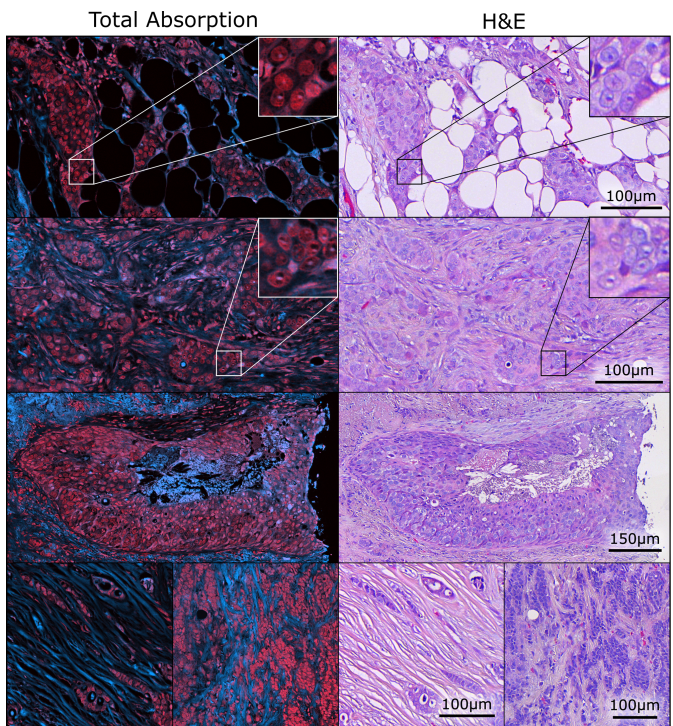

Figure 8: One to one comparison of PARS total absorption (TA) images with the gold standard H&E stain of a breast needle core biopsy. In the colored TA image, the radiative contrast is blue and the non-radiative is red.

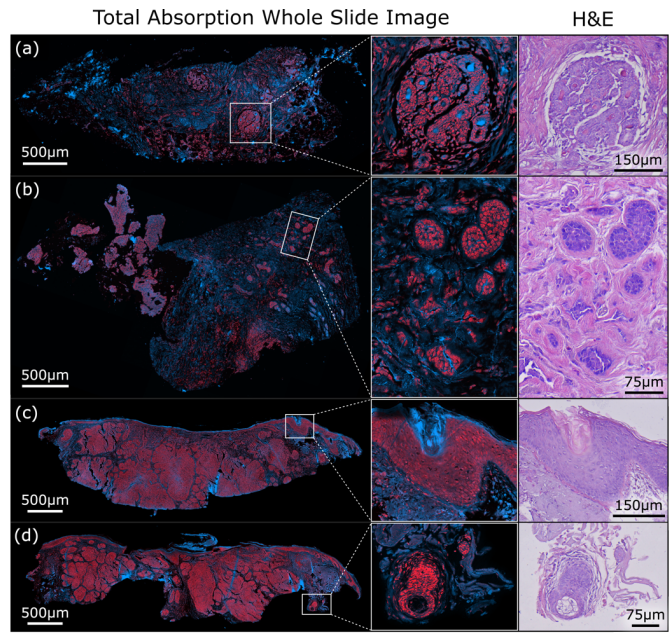

Figure 9: Four whole slide PARS TA images presented at low and high magnifications with corresponding high magnification H&E images. Parts (a) and (b) show two whole slide images of malignant human breast tissue. Parts (c) and (d) show three PARS whole slide images of malignant human skin tissue.

Next, whole slide TA visualizations were produced to demonstrate their potential as a stain-free alternative to H&E whole slide images. Figure 9 shows a set of four TA whole slide images with corresponding one-to-one H&E at high magnification. Pathologists are able to analyze these label-free slides at the same magnifications, using the same or similar digital pathology tools as they would with gold standard H&E stained slides. Figure 9(a) and (b) show whole slide sections of human breast tissue exhibiting invasive mucinous carcinoma and ductal carcinoma, respectively. At both low and high magnifications, irregular and poorly organised ductal structures can be seen in the breast tissue samples, indicative of a pathologic process. Figure 9(c) and (d) show whole slide sections of human skin tissue exhibiting basal cell carcinoma. At low magnifications, the tumour nodules are easily identifiable in both skin tissue samples as dense nests of basaloid cells primarily below the skin surface with some extending from the epidermis. Of note, the TA colorization heatmap strongly corresponds to and facilitates identifying abnormal areas quickly in comparison to standard H&E images. For reference, the normal epidermis of the skin can be seen at high magnification in Figure 9(c) along with high resolution details of cell arrangements, nuclei and subnuclear structures. The stratum cornea can be seen primarily with the blue radiative contrast. Furthermore, the high magnification view of Figure 9(d) shows a sebaceous gland in the tissue sample.

### D. Whole Slide Virtual H&E Staining

Pathologists are trained to analyze tissue sections stained in the style of H&E, and while the whole slide TA images in figure 9 show analogous contrast, emulation of H&E staining style can facilitate a better interpretation experience. Furthermore, machine learning techniques for disease classification and image segmentation have mostly been focused on H&E staining styles [54]. Recently our group has shown such H&E emulation by using the generative machine learning model pix2pix to intelligently combine the radiative, non-radiative and scattering contrast channels [51]. With the automated whole slide scanning system and raw data quality improvements presented here, we are now able to virtually stain whole slide images with H&E. A section of malignant breast tissue, shown in Figure 10(b), was scanned, and then virtually stained to resemble standard H&E. The virtual H&E model for this sample was trained on a separate whole slide image of breast tissue. At low magnifications we can see an almost identical whole slide image and are able to make out the same ductal and lobular structures as well as areas of fatty tissue. At high magnifications the virtual H&E staining compares very well to the gold standard, as determined by our clinical colleagues.

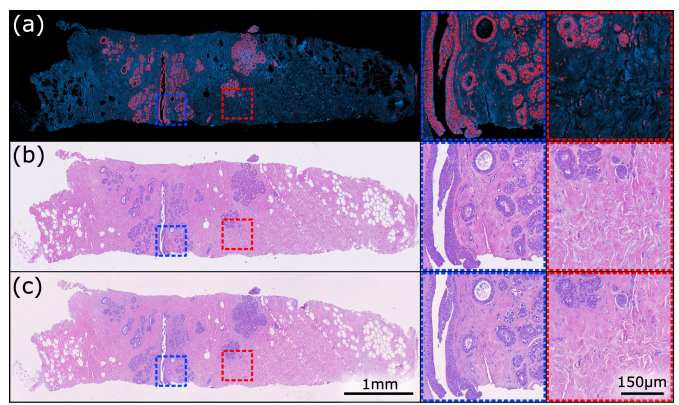

Figure 10: Comparison between the PARS TA, virtual H&E, and corresponding ground truth H&E whole slide images of a breast needle core biopsy. (a) Shows a raw TA whole slide image. (b) Shows the PARS TA whole slide image virtually stained with H&E. (c) Shows the ground truth H&E whole slide image. Two example areas are shown in higher magnification.

## IV. CONCLUSION

Here we present a PARS microscope optimized for label-free histology imaging. We have shown the first transmission mode PARS architecture with optics customized for scanning of thin transmissible samples. The system performs label-free automated whole slide scanning using the autofocus workflow and contrast leveling algorithms presented here. Whole slide images of all three contrast channels (scattering, radiative and non-radiative) were shown in malignant human breast and skin tissue samples. Slides were viewable at magnifications up to 40x and recovered high-resolution subcellular diagnostic details. Clinically relevant features were identified in both the radiative and non-radiative contrast channels as well as in the combined total-absorption (TA) whole slide images. We demonstrated the close correspondence and analogous contrast of the label-free TA images to gold standard H&E staining. Our previously reported pix2pix virtual staining model was successfully applied to an entire breast tissue section, highlighting the system's potential for label-free whole slide H&E emulation. We demonstrated that PARS imaging is non-destructive, permitting standard diagnostic quality H&E staining to be performed sequentially. This ability to capture and virtually stain one-to-one whole slide PARS images is crucial for ongoing blinded clinical pathology validation studies comparing PARS virtual H&E images to ground truth H&E. In future work we aim to capture additional radiative and non-radiative contrasts for improved molecular specificity and emulation of multiple specialized stains. This work represents an important step towards an all-digital histopathologic system to revolutionize the field of histology by overcoming the limitations of century-old tissue processing and staining techniques.

**Acknowledgements**
The authors thank Dr. Ally-Khan Somani, Dr. Gilbert Bigras and the Cross-Cancer Institute in Edmonton, Alberta for providing human breast and skin tissue samples. The authors also thank Hager Gaouda for helping prepare and stain the tissue samples used in this study.

The authors thank the following sources for funding used during this project. Natural Sciences and Engineering Research Council of Canada (DGECR-2019-00143, RGPIN2019-06134); Canada Foundation for Innovation (JELF #38000); Mitacs Accelerate (IT13594); University of Waterloo Startup funds; Centre for Bioengineering and Biotechnology (CBB Seed fund); illumiSonics Inc (SRA #083181); New frontiers in research fund – exploration (NFRFE-2019-01012); The Canadian Institutes of Health Research (CIHR PJT 185984).

**Author Contribution Statement**
J.T. designed and implemented the whole slide scanning framework, conducted experiments, collected TA whole side scans, prepared the figures, and wrote the main manuscript. B.R.E helped with the scanning framework, contrast leveling algorithm and helped scan TA samples. Both J.T. and B.R.E implemented the transmission mode PARS architecture. M.B. performed the pix2pix virtual H&E whole slide colourization. D.D. and J.R.M. prepared and collected tissue specimens and provided clinical feedback and consultation in the assessment of the results. P.H.R directed and organized the project and oversaw experimental work and manuscript writing as the principal investigator.

**Competing Interests**
Authors James Tweel, Benjamin Ecclestone, Deepak Dinakaran, John R. Mackey, and Parsin Haji Reza all have financial interests in IllumiSonics which has provided funding to the PhotoMedicine Labs. Author Marian Boktor does not have any competing interests.

**References**
[1] C. E. Day, Ed., *Histopathology: Methods and Protocols*, vol. 1180. in Methods in Molecular Biology, vol. 1180. New York, NY: Springer, 2014. doi: 10.1007/978-1-4939-1050-2.
[2] T. S. Gurina and L. Simms, "Histology, Staining," in *StatPearls*, Treasure Island (FL): StatPearls Publishing, 2023. Accessed: Mar. 24, 2023. [Online]. Available: http://www.ncbi.nlm.nih.gov/books/NBK557663/
[3] J. S. Makki, "Diagnostic Implication and Clinical Relevance of Ancillary Techniques in Clinical Pathology Practice," *Clin. Med. Insights Pathol.*, vol. 9, pp. 5–11, Mar. 2016, doi: 10.4137/CPath.S32784.
[4] A. H. Fischer, K. A. Jacobson, J. Rose, and R. Zeller, "Hematoxylin and eosin staining of tissue and cell sections," *CSH Protoc.*, vol. 2008, p. pdb.prot4986, May 2008, doi: 10.1101/pdb.prot4986.
[5] L. E. Morrison, M. R. Lefever, H. N. Lewis, M. J. Kapadia, and D. R. Bauer, "Conventional histological and cytological staining with simultaneous immunohistochemistry enabled by invisible chromogens," *Lab. Invest.*, vol. 102, no. 5, Art. no. 5, May 2022, doi: 10.1038/s41374-021-00714-2.
[6] L. Brown, "Improving histopathology turnaround time: a process management approach," *Curr. Diagn. Pathol.*, vol. 10, no. 6, pp. 444–452, Dec. 2004, doi: 10.1016/j.cdip.2004.07.008.

[7] L. Pantanowitz *et al.*, "Review of the current state of whole slide imaging in pathology," *J. Pathol. Inform.*, vol. 2, p. 36, 2011, doi: 10.4103/2153-3539.83746.

[8] F. Großerueschkamp, H. Jütte, K. Gerwert, and A. Tannapfel, "Advances in Digital Pathology: From Artificial Intelligence to Label-Free Imaging," *Visc. Med.*, vol. 37, no. 6, pp. 482–490, Dec. 2021, doi: 10.1159/000518494.

[9] V. Marx, "It's free imaging — label-free, that is," *Nat. Methods*, vol. 16, no. 12, Art. no. 12, Dec. 2019, doi: 10.1038/s41592-019-0664-8.

[10] F. G. Bechara *et al.*, "Histomorphologic correlation with routine histology and optical coherence tomography," *Skin Res. Technol. Off. J. Int. Soc. Bioeng. Skin ISBS Int. Soc. Digit. Imaging Skin ISDIS Int. Soc. Skin Imaging ISSI*, vol. 10, no. 3, pp. 169–173, Aug. 2004, doi: 10.1111/j.1600-0846.2004.00038.x.

[11] T. Maier, M. Braun-Falco, T. Hinz, M. H. Schmid-Wendtner, T. Ruzicka, and C. Berking, "Morphology of basal cell carcinoma in high definition optical coherence tomography: en-face and slice imaging mode, and comparison with histology," *J. Eur. Acad. Dermatol. Venereol. JEADV*, vol. 27, no. 1, pp. e97-104, Jan. 2013, doi: 10.1111/j.1468-3083.2012.04551.x.

[12] T. Gambichler *et al.*, "Comparison of histometric data obtained by optical coherence tomography and routine histology," *J. Biomed. Opt.*, vol. 10, no. 4, p. 44008, 2005, doi: 10.1117/1.2039086.

[13] L. P. Hariri, M. Mino-Kenudson, M. Lanuti, A. J. Miller, E. J. Mark, and M. J. Suter, "Diagnosing lung carcinomas with optical coherence tomography," *Ann. Am. Thorac. Soc.*, vol. 12, no. 2, pp. 193–201, Feb. 2015, doi: 10.1513/AnnalsATS.201408-370OC.

[14] G. Popescu, "Quantitative Phase Imaging of Cells and Tissues," *AccessEngineering | McGraw-Hill Education - Access Engineering*, 2011. https://www.accessengineeringlibrary.com/content/book/9780071663427 (accessed Mar. 24, 2023).

[15] Y. N. Nygate *et al.*, "Holographic virtual staining of individual biological cells," *Proc. Natl. Acad. Sci.*, vol. 117, no. 17, pp. 9223–9231, Apr. 2020, doi: 10.1073/pnas.1919569117.

[16] Y. Rivenson, T. Liu, Z. Wei, Y. Zhang, K. de Haan, and A. Ozcan, "PhaseStain: the digital staining of label-free quantitative phase microscopy images using deep learning," *Light Sci. Appl.*, vol. 8, no. 1, Art. no. 1, Feb. 2019, doi: 10.1038/s41377-019-0129-y.

[17] Z. Wang, K. Tangella, A. Balla, and G. Popescu, "Tissue refractive index as marker of disease," *J. Biomed. Opt.*, vol. 16, no. 11, p. 116017, Nov. 2011, doi: 10.1117/1.3656732.

[18] S. L. Jacques, "Optical properties of biological tissues: a review," *Phys. Med. Biol.*, vol. 58, no. 11, pp. R37-61, Jun. 2013, doi: 10.1088/0031-9155/58/11/R37.

[19] J. Xia, J. Yao, and L. V. Wang, "Photoacoustic tomography: principles and advances," *Electromagn. Waves Camb. Mass*, vol. 147, pp. 1–22, 2014.

[20] F. Jamme *et al.*, "Deep UV autofluorescence microscopy for cell biology and tissue histology," *Biol. Cell*, vol. 105, no. 7, pp. 277–288, 2013, doi: 10.1111/boc.201200075.

[21] Y. Rivenson *et al.*, "Virtual histological staining of unlabelled tissue-autofluorescence images via deep learning," *Nat. Biomed. Eng.*, vol. 3, no. 6, Art. no. 6, Jun. 2019, doi: 10.1038/s41551-019-0362-y.

[22] Y. Zhang, K. de Haan, Y. Rivenson, J. Li, A. Delis, and A. Ozcan, "Digital synthesis of histological stains using micro-structured and multiplexed virtual staining of label-free tissue," *Light Sci. Appl.*, vol. 9, no. 1, Art. no. 1, May 2020, doi: 10.1038/s41377-020-0315-y.

[23] N. H. Patterson *et al.*, "Autofluorescence microscopy as a label-free tool for renal histology and glomerular segmentation." bioRxiv, p. 2021.07.16.452703, Jul. 18, 2021. doi: 10.1101/2021.07.16.452703.

[24] A. C. Croce and G. Bottiroli, "Autofluorescence Spectroscopy and Imaging: A Tool for Biomedical Research and Diagnosis," *Eur. J. Histochem. EJH*, vol. 58, no. 4, p. 2461, Dec. 2014, doi: 10.4081/ejh.2014.2461.

[25] M. Monici, "Cell and tissue autofluorescence research and diagnostic applications," in *Biotechnology Annual Review*, Elsevier, 2005, pp. 227–256. doi: 10.1016/S1387-2656(05)11007-2.

[26] Y.-C. Chen, X. Tan, Q. Sun, Q. Chen, W. Wang, and X. Fan, "Laser-emission imaging of nuclear biomarkers for high-contrast cancer screening and immunodiagnosis," *Nat. Biomed. Eng.*, vol. 1, no. 9, Art. no. 9, Sep. 2017, doi: 10.1038/s41551-017-0128-3.

[27] Y. K. Tao *et al.*, "Assessment of breast pathologies using nonlinear microscopy," *Proc. Natl. Acad. Sci.*, vol. 111, no. 43, pp. 15304–15309, Oct. 2014, doi: 10.1073/pnas.1416955111.

[28] S. Witte *et al.*, "Label-free live brain imaging and targeted patching with third-harmonic generation microscopy," *Proc. Natl. Acad. Sci. U. S. A.*, vol. 108, no. 15, pp. 5970–5975, Apr. 2011, doi: 10.1073/pnas.1018743108.

[29] M. Ji *et al.*, "Rapid, label-free detection of brain tumors with stimulated Raman scattering microscopy," *Sci. Transl. Med.*, vol. 5, no. 201, p. 201ra119, Sep. 2013, doi: 10.1126/scitranslmed.3005954.

[30] F.-K. Lu *et al.*, "Label-free DNA imaging in vivo with stimulated Raman scattering microscopy," *Proc. Natl. Acad. Sci.*, vol. 112, no. 37, pp. 11624–11629, Sep. 2015, doi: 10.1073/pnas.1515121112.

[31] D. A. Orringer *et al.*, "Rapid intraoperative histology of unprocessed surgical specimens via fibre-laser-based stimulated Raman scattering microscopy," *Nat. Biomed. Eng.*, vol. 1, no. 2, Art. no. 2, Feb. 2017, doi: 10.1038/s41551-016-0027.

[32] H. Tu *et al.*, "Stain-free histopathology by programmable supercontinuum pulses," *Nat. Photonics*, vol. 10, no. 8, Art. no. 8, Aug. 2016, doi: 10.1038/nphoton.2016.94.

[33] S. You *et al.*, "Intravital imaging by simultaneous label-free autofluorescence-multiharmonic microscopy," *Nat. Commun.*, vol. 9, no. 1, Art. no. 1, May 2018, doi: 10.1038/s41467-018-04470-8.

[34] T. Kobayashi, K. Nakata, I. Yajima, M. Kato, and H. Tsurui, "Label-Free Imaging of Melanoma with Confocal Photothermal Microscopy: Differentiation between Malignant and Benign Tissue," *Bioengineering*, vol. 5, no. 3, p. 67, Aug. 2018, doi: 10.3390/bioengineering5030067.

[35] M. Schnell et al., "All-digital histopathology by infrared-optical hybrid microscopy," *Proc. Natl. Acad. Sci. U. S. A.*, vol. 117, no. 7, pp. 3388–3396, Feb. 2020, doi: 10.1073/pnas.1912400117.

[36] R. Cao et al., "Label-free intraoperative histology of bone tissue via deep-learning-assisted ultraviolet photoacoustic microscopy," *Nat. Biomed. Eng.*, vol. 7, no. 2, Art. no. 2, Feb. 2023, doi: 10.1038/s41551-022-00940-z.

[37] D.-K. Yao, R. Chen, K. Maslov, Q. Zhou, and L. V. Wang, "Optimal ultraviolet wavelength for in vivo photoacoustic imaging of cell nuclei," *J. Biomed. Opt.*, vol. 17, no. 5, p. 056004, May 2012, doi: 10.1117/1.JBO.17.5.056004.

[38] C. Zhang, Y. S. Zhang, D.-K. Yao, Y. Xia, and L. V. Wang, "Label-free photoacoustic microscopy of cytochromes," *J. Biomed. Opt.*, vol. 18, no. 2, p. 020504, Feb. 2013, doi: 10.1117/1.JBO.18.2.020504.

[39] N. J. M. Haven, K. L. Bell, P. Kedarisetti, J. D. Lewis, and R. J. Zemp, "Ultraviolet photoacoustic remote sensing microscopy," *Opt. Lett.*, vol. 44, no. 14, pp. 3586–3589, Jul. 2019, doi: 10.1364/OL.44.003586.

[40] S. Abbasi et al., "All-optical Reflection-mode Microscopic Histology of Unstained Human Tissues," *Sci. Rep.*, vol. 9, no. 1, p. 13392, Sep. 2019, doi: 10.1038/s41598-019-49849-9.

[41] P. Hajireza, W. Shi, K. Bell, R. J. Paproski, and R. J. Zemp, "Non-interferometric photoacoustic remote sensing microscopy," *Light Sci. Appl.*, vol. 6, no. 6, Art. no. 6, Jun. 2017, doi: 10.1038/lsa.2016.278.

[42] B. S. Restall, N. J. M. Haven, P. Kedarisetti, and R. J. Zemp, "In vivo combined virtual histology and vascular imaging with dual-wavelength photoacoustic remote sensing microscopy," *OSA Contin.*, vol. 3, no. 10, pp. 2680–2689, Oct. 2020, doi: 10.1364/OSAC.398269.

[43] Z. Hosseinaee, Nima Abbasi, N. Pellegrino, L. Khalili, L. Mukhangaliyeva, and P. Haji Reza, "Functional and structural ophthalmic imaging using noncontact multimodal photoacoustic remote sensing microscopy and optical coherence tomography," *Sci. Rep.*, vol. 11, no. 1, Art. no. 1, Jun. 2021, doi: 10.1038/s41598-021-90776-5.

[44] Z. Hosseinaee et al., "Functional photoacoustic remote sensing microscopy using a stabilized temperature-regulated stimulated Raman scattering light source," *Opt. Express*, vol. 29, no. 19, pp. 29745–29754, Sep. 2021, doi: 10.1364/OE.434004.

[45] L. Mukhangaliyeva et al., "Deformable mirror-based photoacoustic remote sensing (PARS) microscopy for depth scanning," *Biomed. Opt. Express*, vol. 13, no. 11, pp. 5643–5653, Nov. 2022, doi: 10.1364/BOE.471770.

[46] K. Bell et al., "Reflection-mode virtual histology using photoacoustic remote sensing microscopy," *Sci. Rep.*, vol. 10, no. 1, p. 19121, Nov. 2020, doi: 10.1038/s41598-020-76155-6.

[47] B. R. Ecclestone et al., "Improving maximal safe brain tumor resection with photoacoustic remote sensing microscopy," *Sci. Rep.*, vol. 10, no. 1, p. 17211, Oct. 2020, doi: 10.1038/s41598-020-74160-3.

[48] K. Bell, L. Mukhangaliyeva, L. Khalili, and P. H. Reza, "Hyperspectral Absorption Microscopy Using Photoacoustic Remote Sensing," *Opt. Express*, vol. 29, no. 15, p. 24338, Jul. 2021, doi: 10.1364/OE.430403.

[49] P. Kedarisetti, N. J. M. Haven, B. S. Restall, M. T. Martell, and R. J. Zemp, "Label-free lipid contrast imaging using non-contact near-infrared photoacoustic remote sensing microscopy," *Opt. Lett.*, vol. 45, no. 16, pp. 4559–4562, Aug. 2020, doi: 10.1364/OL.397614.

[50] B. R. Ecclestone, K. Bell, S. Sparkes, D. Dinakaran, J. R. Mackey, and P. Haji Reza, "Label-free complete absorption microscopy using second generation photoacoustic remote sensing," *Sci. Rep.*, vol. 12, no. 1, Art. no. 1, May 2022, doi: 10.1038/s41598-022-11235-3.

[51] M. Boktor et al., "Virtual histological staining of label-free total absorption photoacoustic remote sensing (TA-PARS)," *Sci. Rep.*, vol. 12, no. 1, Art. no. 1, Jun. 2022, doi: 10.1038/s41598-022-14042-y.

[52] T. Yeo, S. Ong, Jayasooriah, and R. Sinniah, "Autofocusing for tissue microscopy," *Image Vis. Comput.*, vol. 11, no. 10, pp. 629–639, Dec. 1993, doi: 10.1016/0262-8856(93)90059-P.

[53] T. L. Sellaro et al., "Relationship between magnification and resolution in digital pathology systems," *J. Pathol. Inform.*, vol. 4, no. 1, p. 21, Jan. 2013, doi: 10.4103/2153-3539.116866.

[54] J. van der Laak, G. Litjens, and F. Ciompi, "Deep learning in histopathology: the path to the clinic," *Nat. Med.*, vol. 27, no. 5, Art. no. 5, May 2021, doi: 10.1038/s41591-021-01343-4.